\documentclass{appolb}
\usepackage{epsfig}

\def\Journal#1#2#3#4{{#1} {\bf #2}, #3 (#4)}

\newcommand{\dNdeta}{$\mathrm{d}N_\mathrm{ch}/\mathrm{d}\eta$}

\begin{document}

\title{Identified-particle production and spectra with the ALICE detector in
  pp and Pb--Pb collisions at the LHC
}
\author{Roberto Preghenella (for the ALICE Collaboration)
  \thanks{Presented at \emph{Strangeness in Quark Matter}, 18--24 September 2011,
    Crakow, Poland}
  \address{Centro Studi e Ricerche e Museo Storico della Fisica ``Enrico Fermi'',
    Rome, Italy\\
    Dipartimento di Fisica dell'Universit\`a and Sezione INFN, Bologna,
    Italy\\
    E-mail: {\tt preghenella@bo.infn.it}}
}
\maketitle
\begin{abstract}
  Thanks to its unique capabilities the ALICE experiment can measure the
  production of identified particles and resonances over a wide momentum range
  both in pp and Pb--Pb collisions at the LHC. In this report,
  particle-identification detectors and techniques, as well as achieved
  performance, are shortly reviewed. The current results on hadron transverse
  momentum spectra measured in pp collisions at $\sqrt{s}$~=~0.9~TeV and
  7~TeV, and in Pb--Pb collisions at $\sqrt{s_{\rm NN}}$~=~2.76~TeV are shown. In
  particular, proton-proton results on particle production yields, spectral
  shapes and particle ratios are presented as a function of the collision
  energy and compared to previous experiments and commonly-used Monte Carlo
  models. Particle spectra, yields and ratios in Pb--Pb are measured as
  a function of the collision centrality and the results are compared with
  published RHIC data in Au--Au collisions at $\sqrt{s_{\rm NN}}$~=~0.2~TeV
  and predictions for the LHC. 
\end{abstract}

\section{Introduction}\label{sec:introduction}
ALICE (A Large Ion Collider Experiment) is a general-purpose heavy-ion
experiment at the CERN LHC (Large Hadron Collider) aimed at studying the
physics of strongly-interacting matter and the quark--gluon plasma. A unique
design has been adopted for the ALICE detector to fulfill tracking and
particle-identification requirements~\cite{ref:ALICEperf}. Thanks to these features the experiment
is able to identify hadrons in a wide momentum range by combining
different detecting systems and techniques, as discussed in 
Section~\ref{sec:pid}. Results on
hadron spectra and yields at mid-rapidity are presented in
Section~\ref{sec:ppresults} for pp collisions at $\sqrt{s}$~=~0.9~TeV and
7~TeV and in Section~\ref{sec:pbpbresults} for Pb--Pb collisions at
$\sqrt{s_{\rm NN}}$~=~2.76~TeV.

\section{Particle identification}\label{sec:pid}

In this section the particle-identification (PID) detectors relevant for the
analyses presented in this paper are briefly discussed, namely the \emph{Inner
  Tracking System} (ITS), 
the \emph{Time-Projection Chamber} (TPC) and the \emph{Time-Of-Flight} detector
(TOF). A detailed review of the ALICE experiment and of its PID capabilities can
be found in~\cite{ref:ALICEperf}. The ITS is a six-layer silicon detector located at radii between 4 and 43
cm. Four of the six layers provide $dE/dx$ measurements and are used for
particle identification in the non-relativistic ($1/\beta^2$)
region. Moreover, using the ITS as a standalone tracker enables one to
reconstruct and identify low-momentum particles (below~200~MeV/c) not reaching the main tracking
systems. The TPC is the main central-barrel tracking detector of ALICE and
provides three-dimensional hit information and specific energy-loss
measurements with up to 159 samples. With the measured particle momentum and
$\langle dE/dx \rangle$ the particle type can be determined by comparing the
measurements against the Bethe-Bloch expectation. The TOF detector is a
large-area array of Multigap Resistive Plate Chambers (MRPC) and covers the central
pseudorapidity region ($\left| \eta \right| <$~0.9, full azimuth). Particle
identification is performed by matching momentum and trajectory-length
measurements performed by the tracking system with the time-of-flight
information provided by the TOF system. The total time-of-flight resolution is
about 85~ps in Pb--Pb collisions (about 120~ps in pp collisions) and it is determined by the time resolution
of the detector itself and by the start-time resolution.

The transverse momentum spectra of primary $\rm \pi^{\pm}$, $\rm K^{\pm}$,
$\rm p$ and
$\rm \bar{p}$ are measured at mid-rapidity ($\left|y\right|~<~0.5$) combining the
techniques and detectors described above. Primary particles
are defined as prompt particles produced in the collision and all decay
products, except products from weak decay of strange particles. The
contribution from the feed-down of weakly-decaying particles to $\rm \pi^{\pm}$,
$\rm p$ and $\rm \bar{p}$ and from protons from material are subtracted by fitting the
data using Monte Carlo templates of the DCA\footnote{Distance of Closest
  Approach to the reconstructed primary vertex.} distributions. Particles can
also be identified in ALICE via their characteristics decay 
topology or invariant mass fits. This, combined with the direct identification
of the decay daughters, allows one to reconstruct weakly-decaying particles and
hadronic resonances with an improved signal-to-background ratio.

\section{Results in pp collisions}\label{sec:ppresults}

\begin{figure}[t]
  \centering
  \begin{minipage}[c]{0.57\linewidth}
    \centering
    \includegraphics[width=\textwidth]{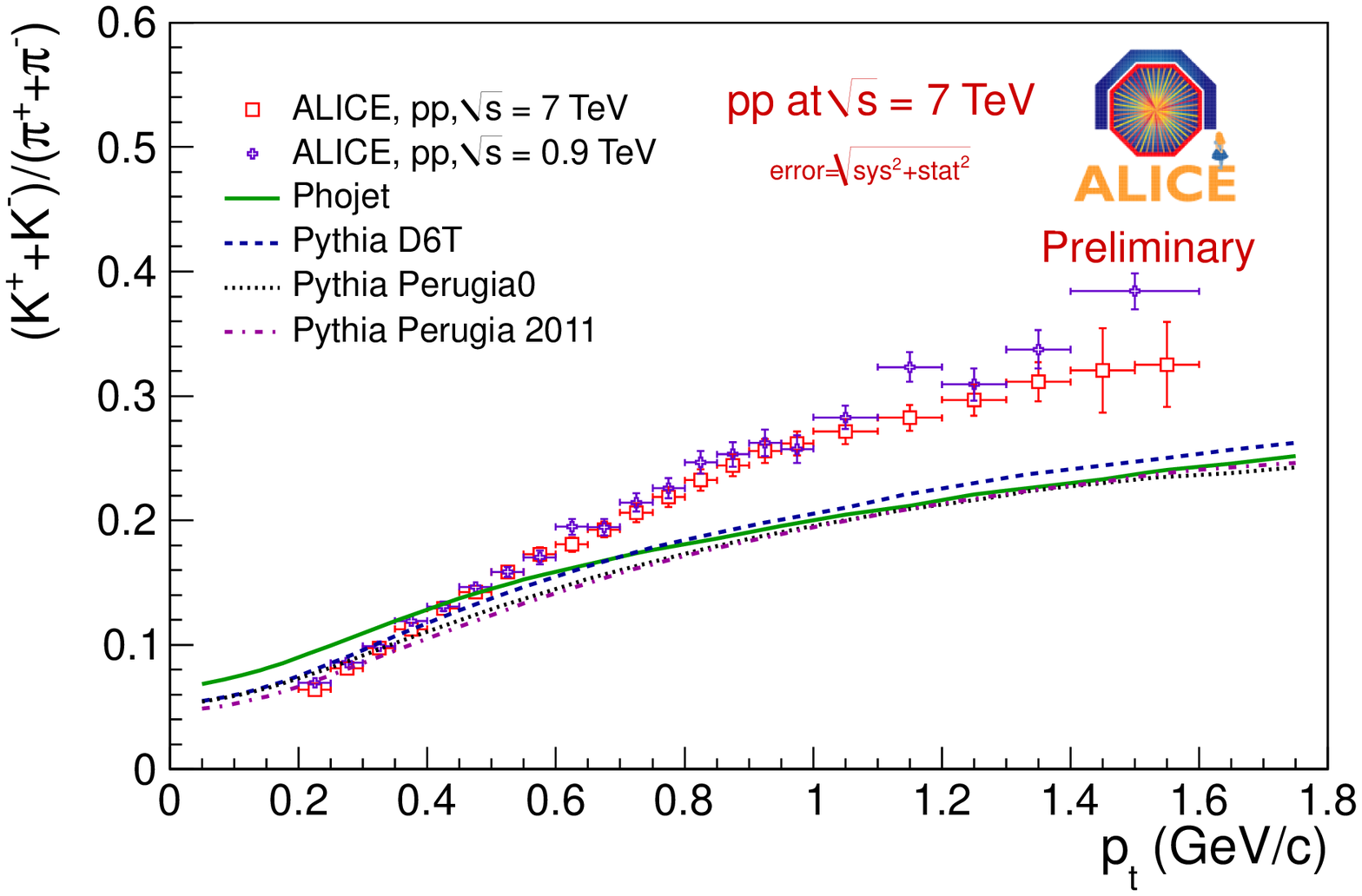}
  \end{minipage}
  \begin{minipage}[c]{0.57\linewidth}
    \centering
    \includegraphics[width=\linewidth]{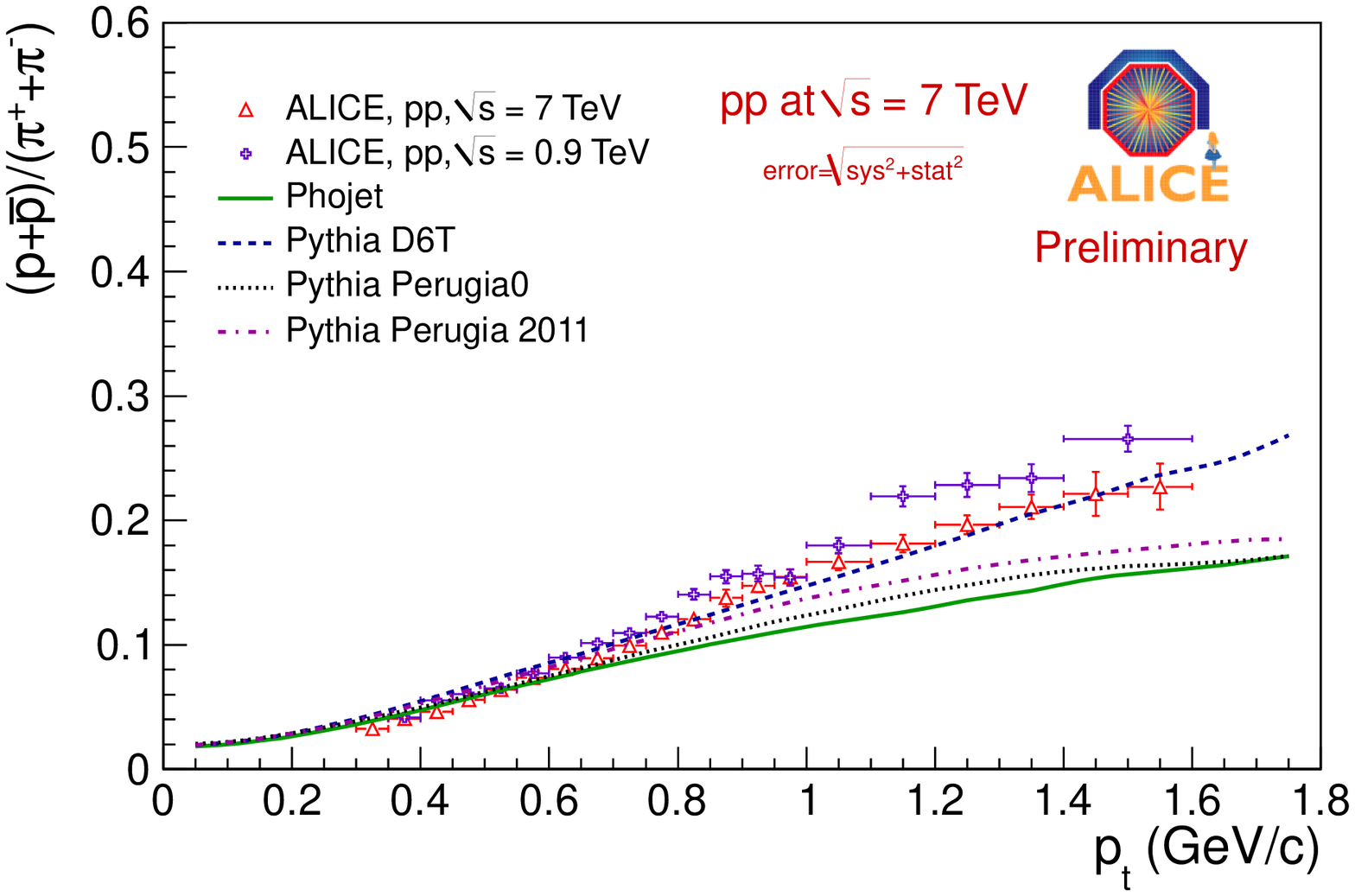}
  \end{minipage}
  \caption{$\rm K/\pi$ (top) and $\rm p/\pi$ (bottom) production ratios as a function
    of $p_{T}$ measured in pp collisions at $\sqrt{s}$~=~0.9~TeV and 7~TeV
    compared to various Monte Carlo models at $\sqrt{s}$~=~7~TeV.}
  \label{fig:ratiopp}
\end{figure}

The transverse momentum spectra of primary $\rm \pi^{\pm}$, $\rm K^{\pm}$, $\rm p$ and
$\rm \bar{p}$ have been measured in minimum-bias pp collisions at
$\sqrt{s}$~=~0.9~TeV and 7~TeV. The measurements performed at
$\sqrt{s}$~=~0.9~TeV and the details of the analysis are already published
in~\cite{ref:spectra900}. The results of the analysis performed at
$\sqrt{s}$~=~7~TeV were reported at this
conference~\cite{ref:BarbaraSQM}. The ratios $\rm K/\pi$ and $\rm p/\pi$ as a function of $p_{T}$ are shown in
Figure~\ref{fig:ratiopp} comparing measurements at $\sqrt{s}$~=~0.9~TeV and
7~TeV. Both ratios do not show evident energy dependence and the same
holds for the $p_{T}$-integrated production ratios which are also observed to be rather
independent of the collision energy from 0.9~TeV to 7~TeV. Moreover, no difference between $\rm p/\pi^{+}$ and
$\rm \bar{p}/\pi^{-}$ ratios is observed leading to a constant value of about
0.05 and vanishing baryon/antibaryon asymmetry at LHC as already reported in~\cite{ref:panos}.
The comparison with Monte Carlo generators shows that the $p_{T}$-dependent $\rm K/\pi$ ratio is
underestimated at high $p_{T}$ by recent PYTHIA tunes. The same
holds for the $\rm p/\pi$ ratio, though a better agreement with the data is observed
for PYTHIA D6T. 

\begin{figure}[t]
  \centering
  \begin{minipage}[c]{0.48\linewidth}
    \centering
    \includegraphics[width=\textwidth]{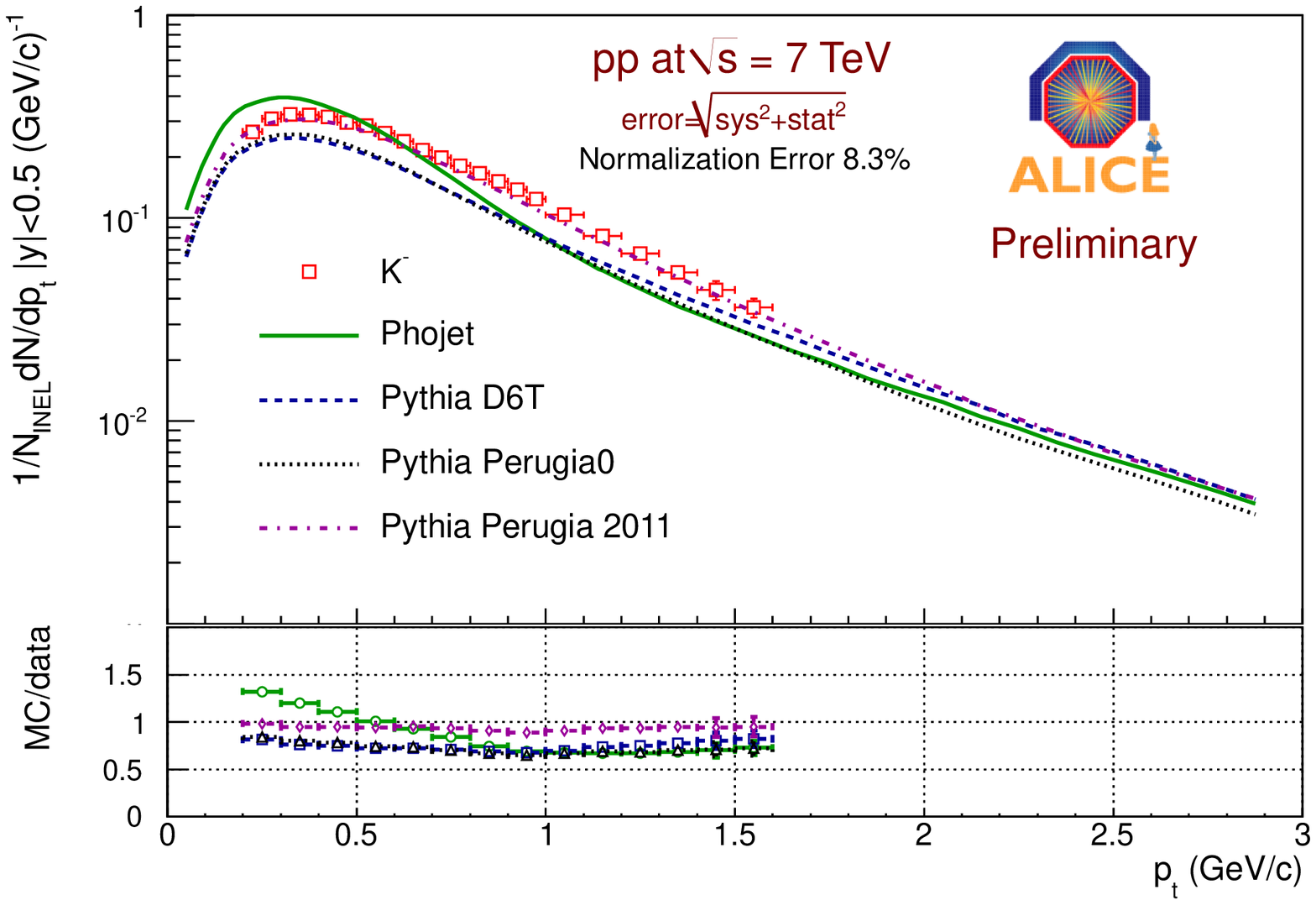}
  \end{minipage}
  \hfill
  \begin{minipage}[c]{0.48\linewidth}
    \centering
    \includegraphics[width=0.9\linewidth]{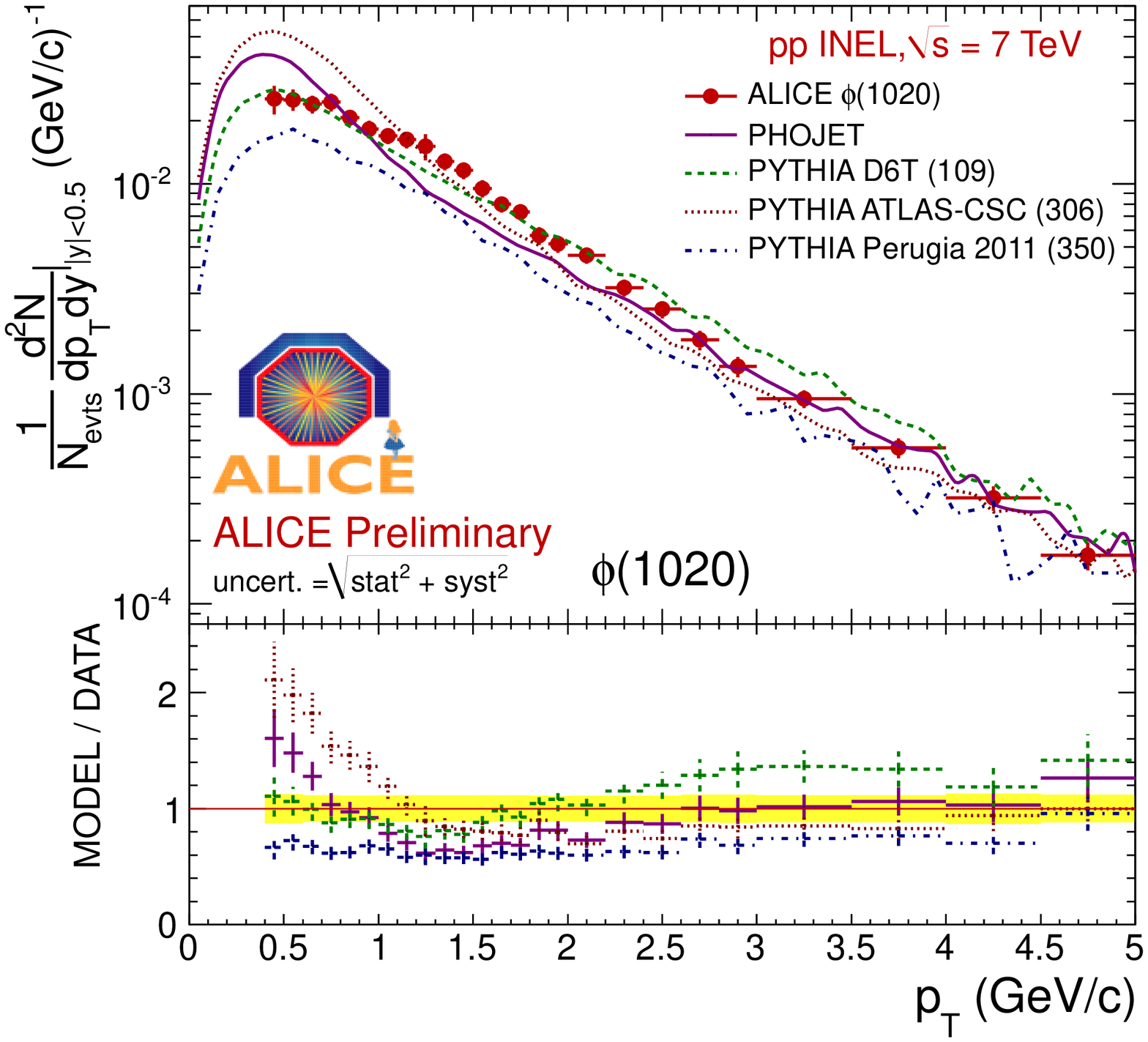}
  \end{minipage}
  \caption{Transverse momentum spectra of $\rm K^{-}$ (left) and $\phi$ (right) in pp collisions at
      $\sqrt{s}$~=~7~TeV compared to various Monte Carlo models.}
  \label{fig:kaonkstarppmc}
\end{figure}

Measurements of the production of multi-strange baryons $\rm \Xi \to \Lambda +
\pi \to p + \pi + \pi$ and $\rm \Omega \to \Lambda + K \to p + \pi + K$ and
of hadronic resonances $\rm K^{*} \to K + \pi$ and $\rm \phi \to K^{+} + K^{-}$ in pp
collisions were also reported at this conference (see~\cite{ref:AntoninSQM}
and~\cite{ref:DhevanSQM}, respectively). When compared to Monte Carlo 
event generators, multi-strange baryon production is under-predicted by various PYTHIA
tunes, though the most recent Perugia-2011 tune shows
an overall better agreement with the data, in particular in the high $p_{T}$ region. $\rm \phi$ resonance
production is rather well described by PHOJET whereas in PYTHIA Perugia-2011
it is slightly under-predicted at low $p_{T}$. It is worth to stress
here that an overall good agreement between the data and PYTHIA Perugia-2011
tune is observed for charged-kaons but not for resonance
production, as can be
deduced from Figure~\ref{fig:kaonkstarppmc}. On the other hand there is still
some work for tuning model for resonances (especially baryons) and
multi-strange baryon production.

\section{Results in Pb--Pb collisions}\label{sec:pbpbresults}

\begin{figure}[tpb]
  \centering
  \begin{minipage}[c]{0.65\linewidth}
    \centering
    \includegraphics[width=\textwidth]{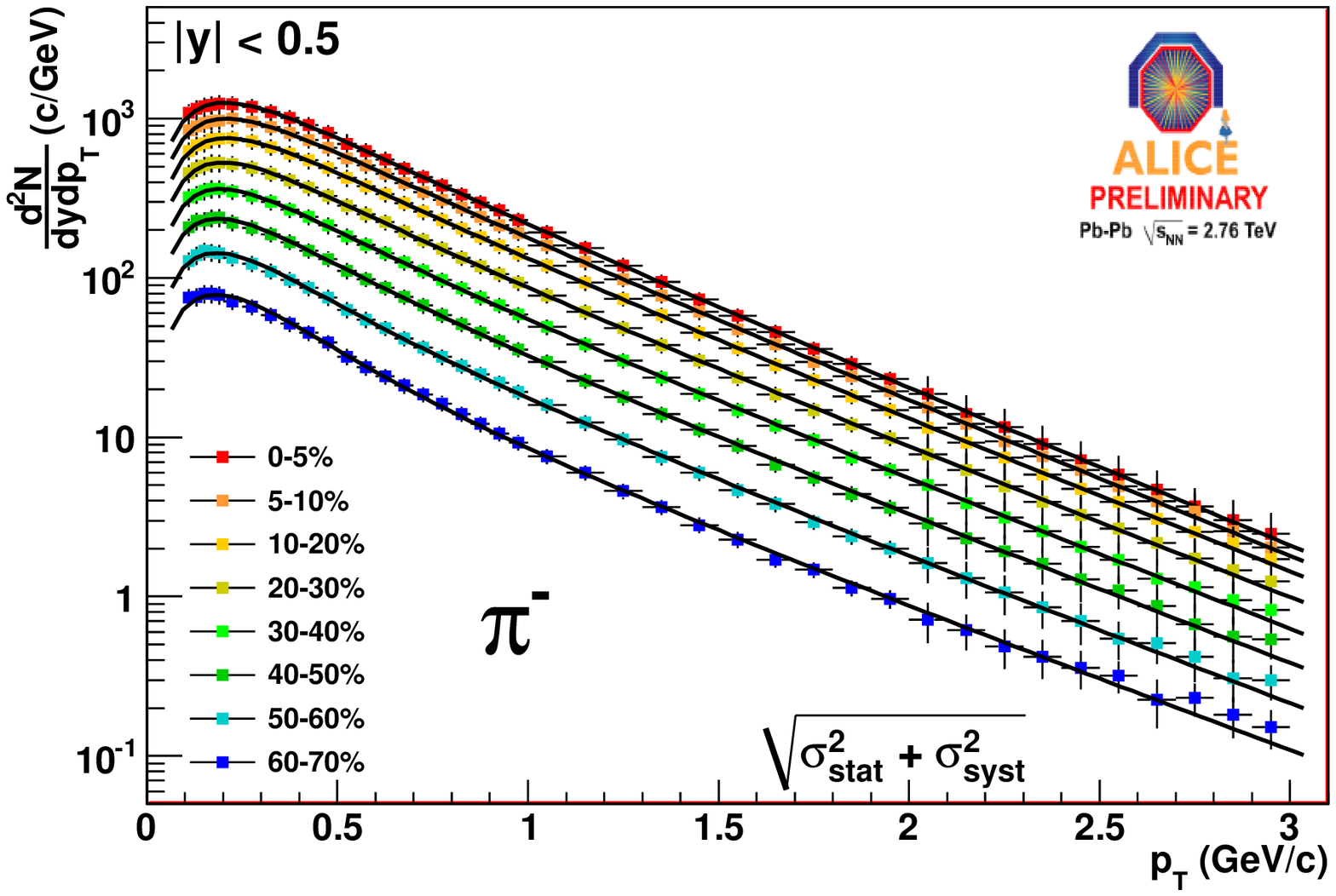}
  \end{minipage}
  \vfill
  \begin{minipage}[c]{0.65\linewidth}
    \centering
    \includegraphics[width=\textwidth]{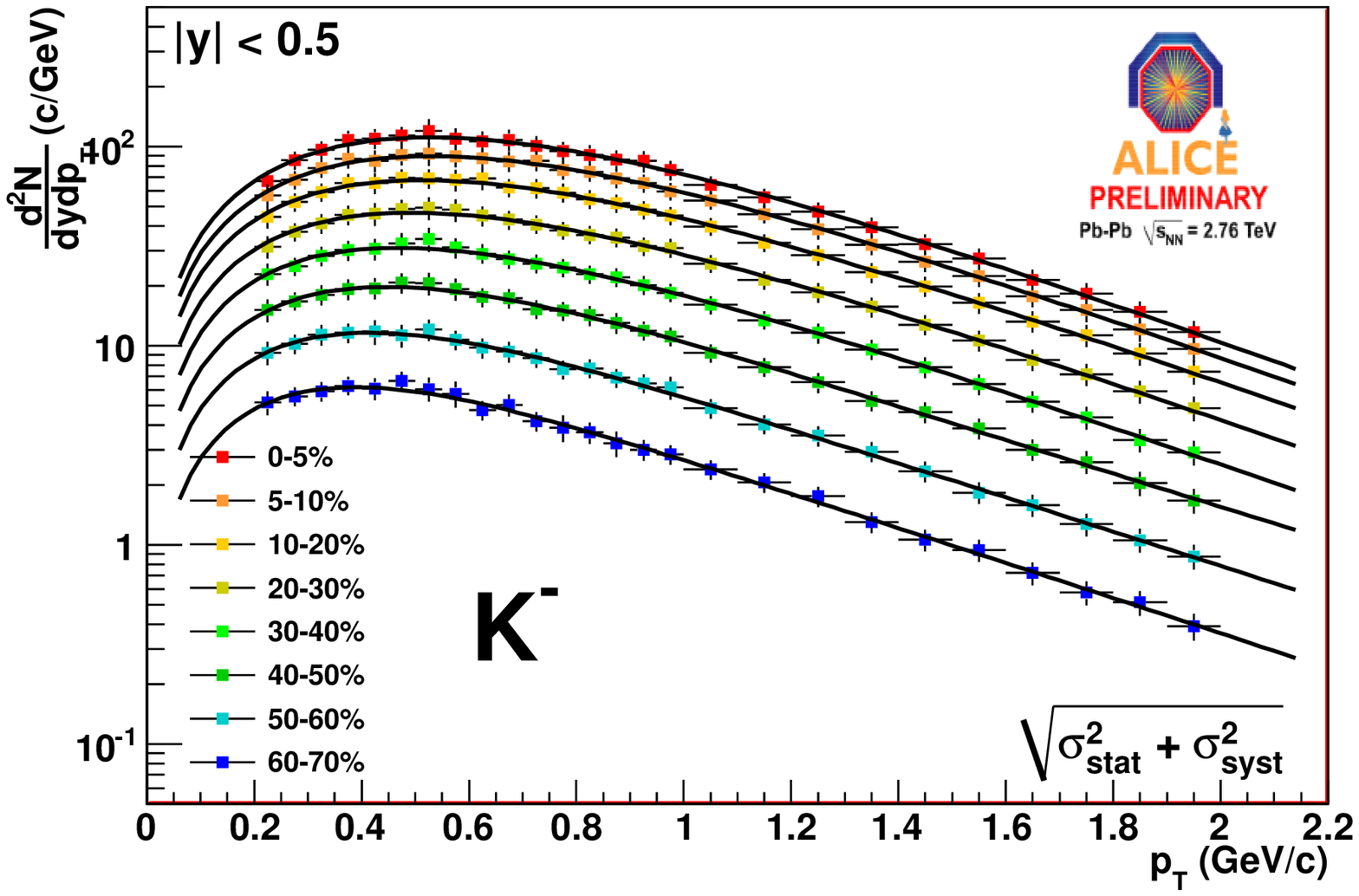}
  \end{minipage}
  \vfill
  \begin{minipage}[c]{0.65\linewidth}
    \centering
    \includegraphics[width=\textwidth]{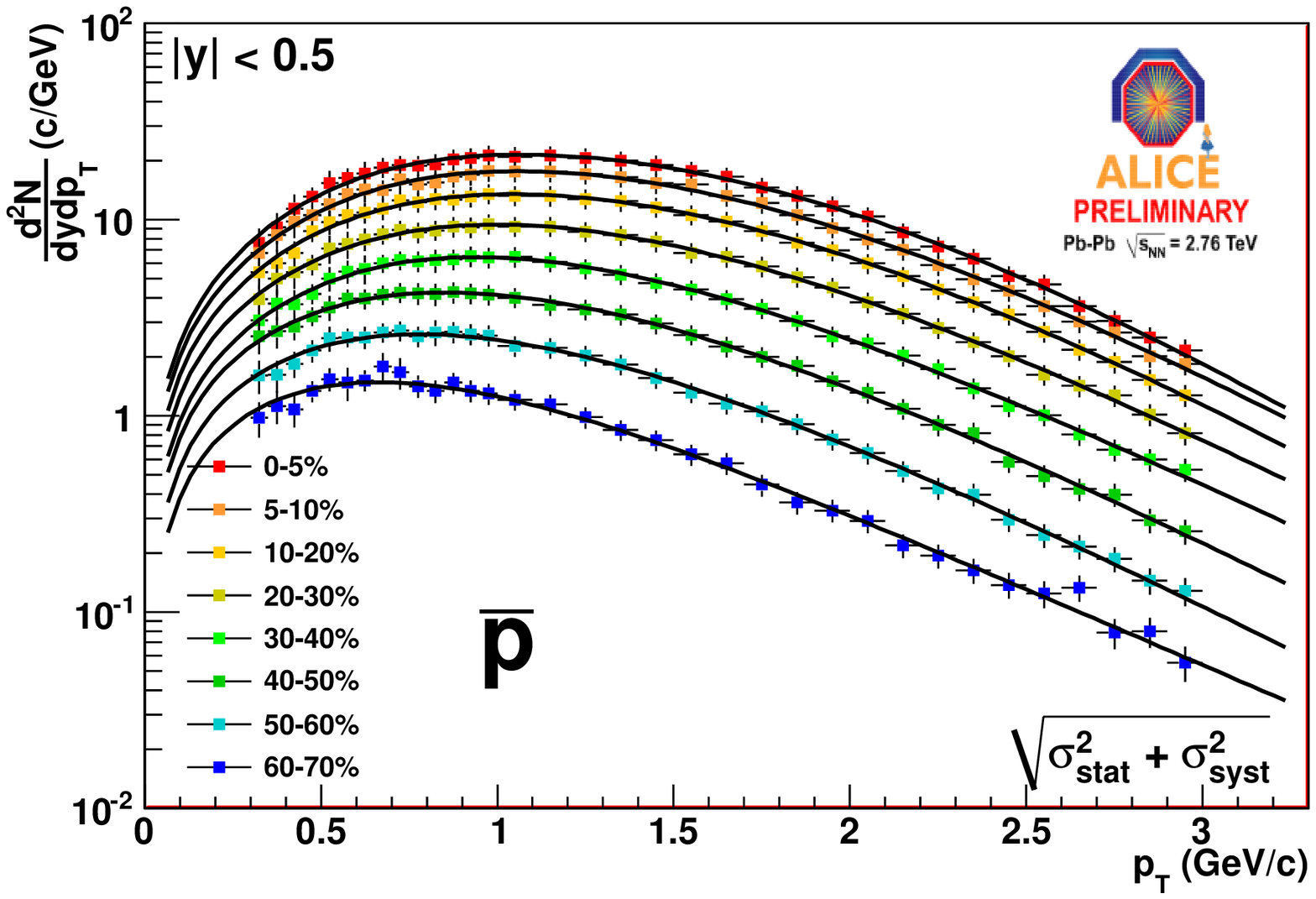}
  \end{minipage}
  \vfill
  \caption{Transverse momentum spectra of primary $\rm \pi^{-}$ (top),
    $\rm K^{-}$ (middle), $\rm \bar{p}$ (bottom) and corresponding fits in Pb--Pb collisions at
    $\sqrt{s_{\rm NN}}$~=~2.76~TeV.}
  \label{fig:pbpbspectra}
\end{figure}

Hadron $p_{T}$ spectra have been
measured in several centrality classes (see~\cite{ref:ALICEpbpb} for details
on centrality selection) from 100~MeV/c up to 3~GeV/c for pions, from
200~MeV/c up to 2~GeV/c for kaons and from 300~MeV/c up to 3~GeV/c for
protons and antiprotons. Individual fits to the data are performed using a
blast-wave parameterization~\cite{ref:blastwave} to extrapolate the
spectra outside the measured $p_{T}$ range. The measured spectra and corresponding fits are shown in
Figure~\ref{fig:pbpbspectra} for primary $\rm \pi^{-}$ (top), $\rm K^{-}$ (middle) and
$\rm \bar{p}$ (bottom). Average transverse 
momenta $\langle p_{T} \rangle$ and integrated 
production yields $dN/dy$ are obtained using the measured data points and the
extrapolation. 

\begin{figure}[t]
  \centering
  \begin{minipage}[c]{0.57\linewidth}
    \centering
    \includegraphics[width=\textwidth]{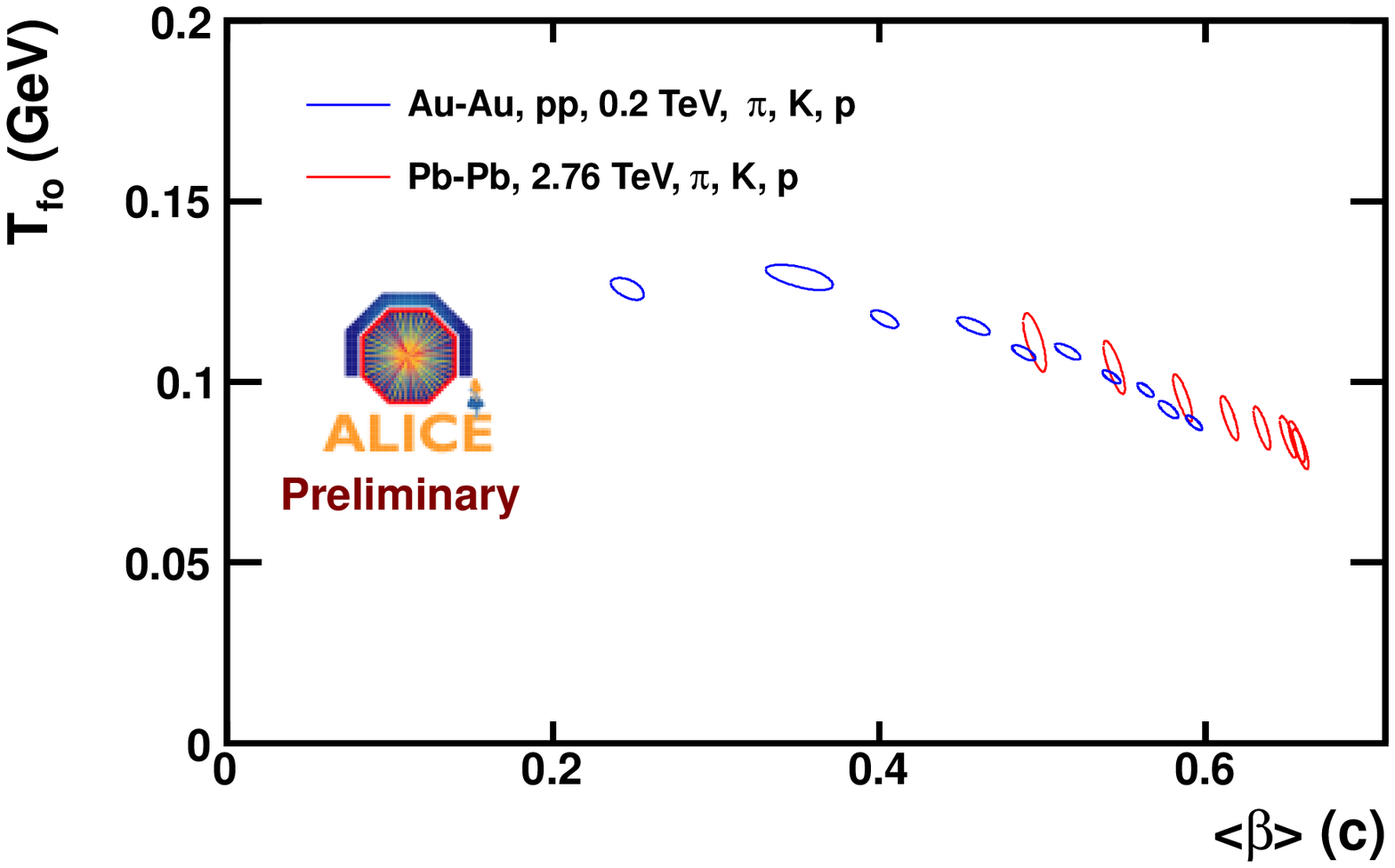}
  \end{minipage}
  \begin{minipage}[c]{0.57\linewidth}
    \centering
    \includegraphics[width=0.9\textwidth]{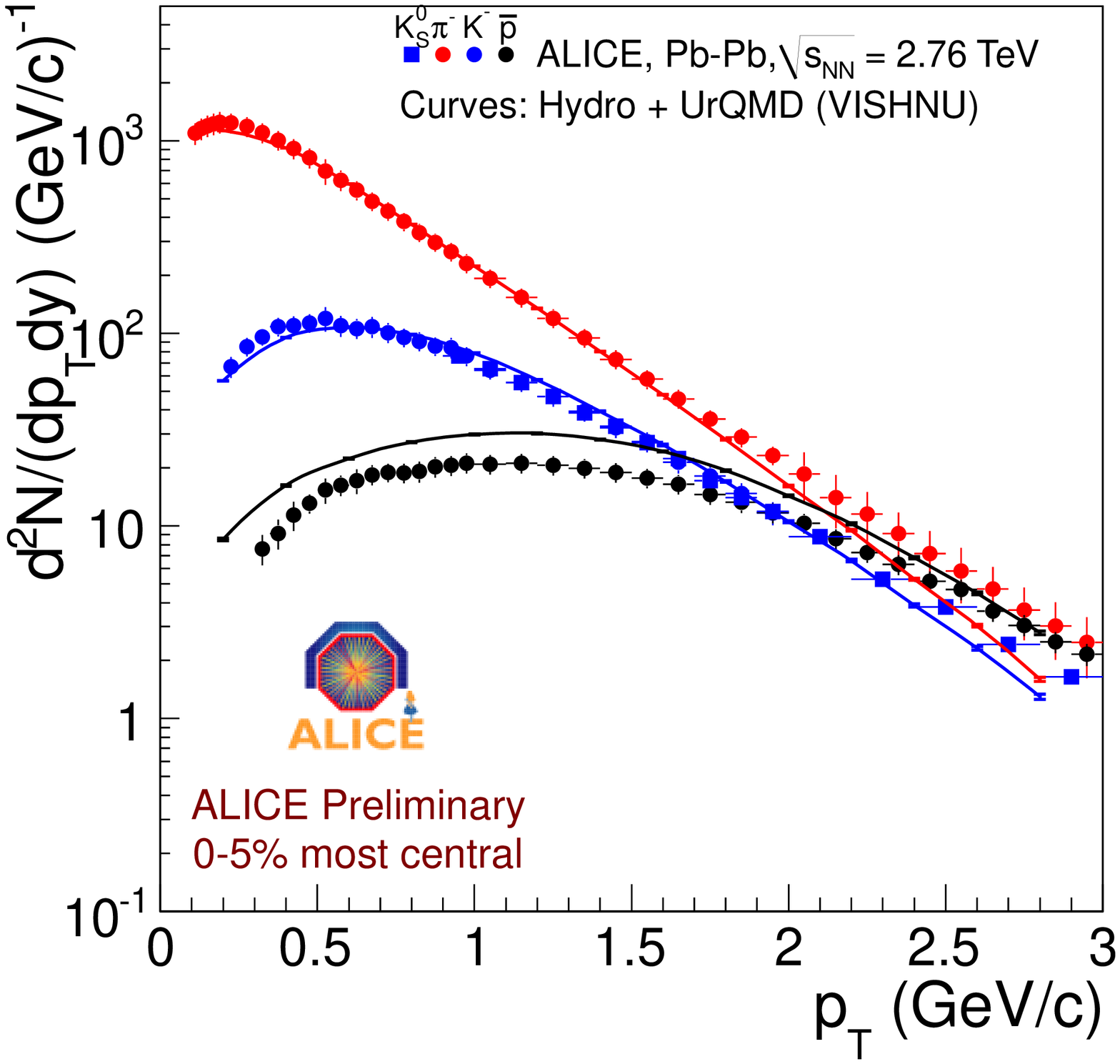}
  \end{minipage}
  \caption{Thermal freeze-out parameters $T_{fo}$ and $\langle \beta \rangle$
    from combined blast-wave fits compared to RHIC data (top). $\rm \pi^{-}$,
    $\rm K^{-}$, $\rm \bar{p}$ spectra in 0-5\% central Pb--Pb 
    collisions compared to hydrodynamic model predictions~\cite{ref:vishnu,ref:vishnuspectra} (bottom).}
  \label{fig:blastwavevishnu}
\end{figure}

As reported in~\cite{ref:BarbaraSQM}, the spectra are observed to be
harder than at RHIC for similar \dNdeta. A detailed study of the
spectral shapes has been done in order to give a quantitative estimate of
the thermal freeze-out temperature $T_{fo}$ and the average transverse flow
$\langle \beta \rangle$. A combined blast-wave fit of the spectra has
been performed in the $p_{T}$ ranges 0.3-1.0~GeV/c, 0.2-1.5~GeV/c and 0.3-3.0~GeV/c
for pions, kaons and protons, respectively. While the $T_{fo}$ parameter is sensitive to the pion fit range because of feed-down of
resonances\footnote{This effect will be investigated in details in the
  future.} the transverse flow $\langle \beta \rangle$ measurement is not,
being dominated by the proton spectral shape. The results obtained on the
thermal freeze-out properties in different centrality bins are compared with
similar measurements performed by the STAR Collaboration at lower energies in
Figure~\ref{fig:blastwavevishnu} (top). A stronger radial flow is observed with respect to
RHIC, being about 10\% larger in the most central collisions at the LHC. The data are also compared to predictions from hydrodynamic models. As already
reported in~\cite{ref:MicheleQM} the pure hydrodynamic
prediction~\cite{ref:purehydro} cannot reproduce the proton shape. A similar
disagreement was observed when comparing proton elliptic flow $v_{2}$ to the
same model~\cite{ref:SnellingsQM}. A new
calculation has been performed using a hybrid 
model which adds a hadronic rescattering and freeze-out stage to the
viscous dynamics~\cite{ref:vishnu}. These new predictions~\cite{ref:vishnuspectra} are compared to the data in
Figure~\ref{fig:blastwavevishnu} (bottom) and the agreement with the proton shape
is better than in a pure hydrodynamic picture. This
suggests that extra flow builds up in the hadronic phase. The
difference in the proton yield can be ascribed to the fact that the model
derives yields from a thermal model with $T_{ch} = 165$~MeV. It is worth to
mention that this model also
reproduces the shape of elliptic flow of identified 
particles as reported at this conference~\cite{ref:NoferoSQM}.

\begin{figure}[t]
  \centering
  \begin{minipage}[c]{0.48\linewidth}
    \centering
    \includegraphics[width=\linewidth]{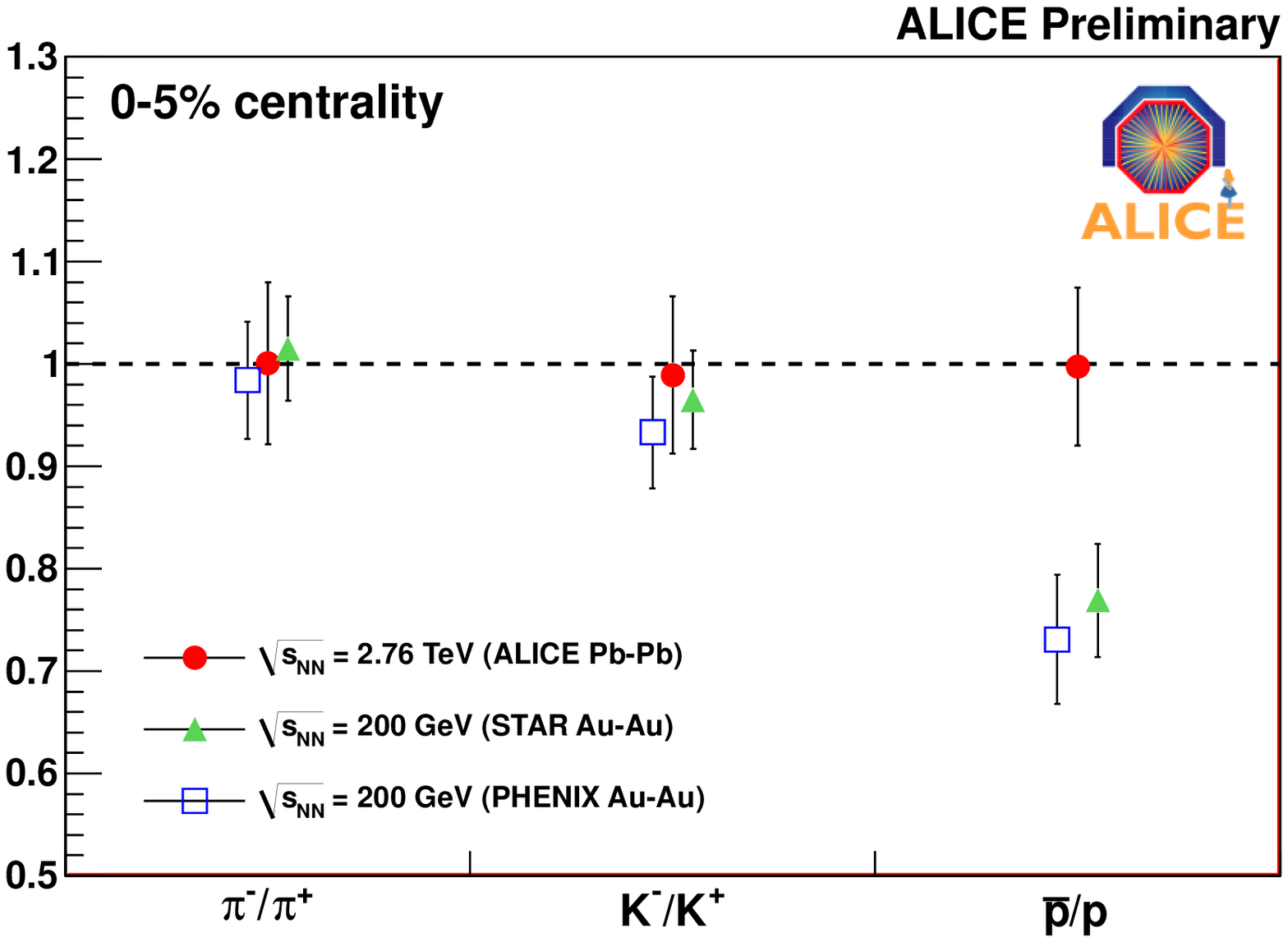}
  \end{minipage}
  \hfill
  \begin{minipage}[c]{0.48\linewidth}
    \centering
    \includegraphics[width=\textwidth]{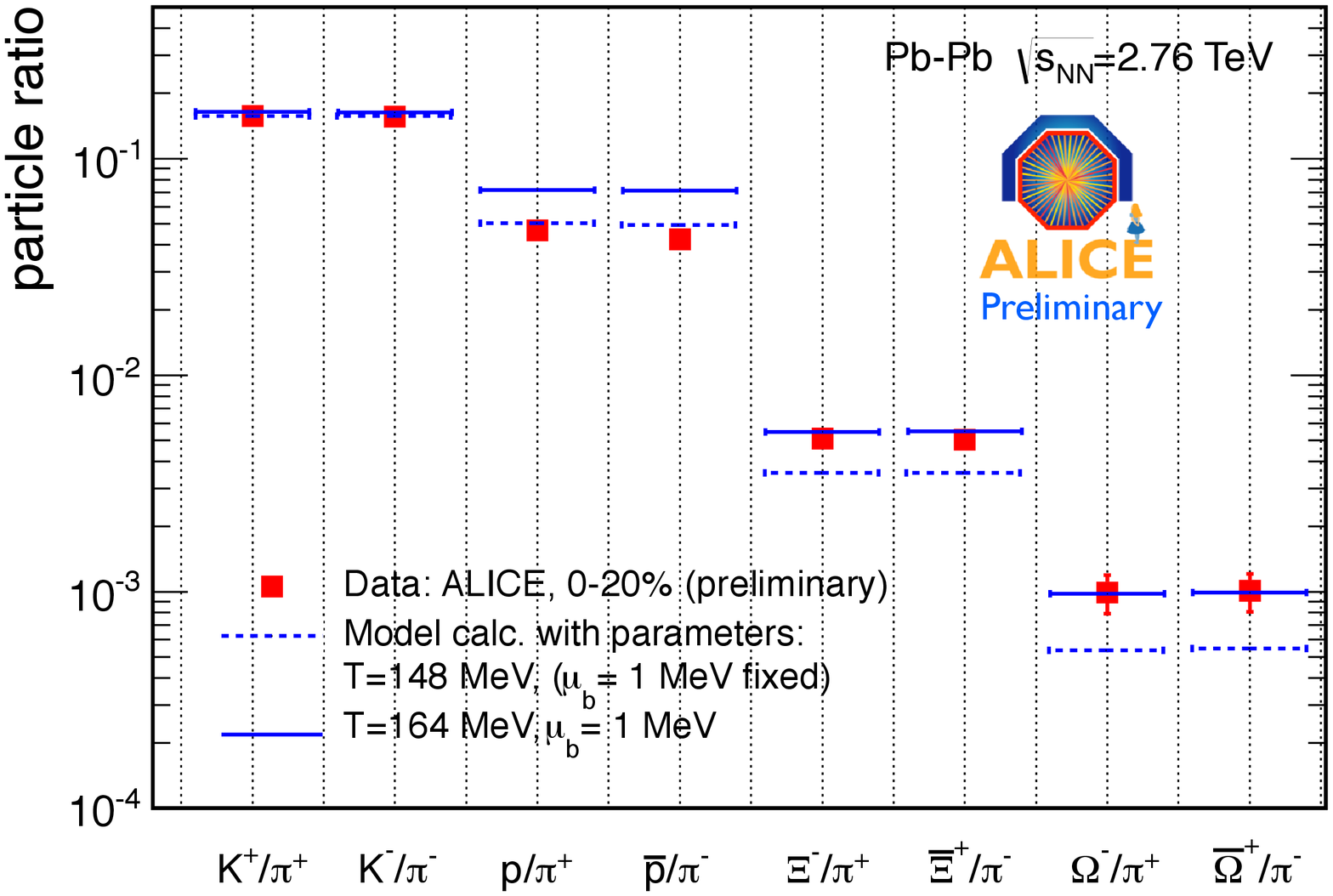}
  \end{minipage}
  \caption{Antiparticle/particle production ratios in the 0-5\% most central Pb--Pb
    collisions~\cite{ref:RHIC} (left). Hadron-production ratios compared to thermal model predictions~\cite{ref:statmodels} (right).}
    \label{fig:mubthermal}
\end{figure}

Antiparticle/particle integrated production ratios are observed to be
consistent with unity for all particle species in all centralities suggesting
that the baryo-chemical potential $\mu_{B}$ is close to zero as expected at LHC
energies. Figure~\ref{fig:mubthermal} (left) compares ALICE results with
RHIC data in Au--Au 
collisions at $\sqrt{s_{\rm NN}}$ =~0.2~TeV~\cite{ref:RHIC} for the 0-5\%
most central collisions. The $p_{T}$-integrated $\rm K^{-}/\pi^{-}$ and
$\rm \bar{p}/\pi^{-}$ ratios and the production of multi-strange baryons in Pb--Pb
collisions were also reported at this
conference (see~\cite{ref:KalweitSQM} and~\cite{ref:MariellaSQM},
respectively). The $\rm K^{-}/\pi^{-}$ production nicely  
follows the trend measured at RHIC. The $\rm \bar{p}/\pi^{-}$ production ratio ($\sim$~0.05) is
significantly lower that the value expected from statistical model predictions
($\sim$~0.07-0.09) with a chemical freeze-out temperature of $T_{ch} =
160-170$~MeV at the LHC~\cite{ref:statmodels}. The measured yields of
several particles normalized to the pion yield in 0-5\% central collisions are
compared to thermal model predictions in Figure~\ref{fig:mubthermal}
(right). With a temperature of $T_{ch} = 164$~MeV  predicted by Andronic
\emph{et al.}~\cite{ref:statmodels} the model reproduces both kaon and
multi-strange baryon
production but overestimates the protons. The same model can be tuned to reproduce
proton yields with an ad-hoc $T_{ch} = 148$~MeV, though multi-strange baryon
production is underestimated in this case.

\section{Conclusions}\label{sec:conclusions}
The tranverse momentum spectra of $\rm \pi^{\pm}$, $\rm K^{\pm}$, $\rm p$ and $\rm \bar{p}$ have
been measured with ALICE in pp collisions at $\sqrt{s}$~=~0.9~TeV and
7~TeV and in Pb--Pb collisions at $\sqrt{s_{\rm NN}}$~=~2.76~TeV,
demonstrating the excellent PID capabilities of the
experiment. Proton-proton results show no evident $\sqrt{s}$ dependence in
hadron production ratios. Monte Carlo generators do not reproduce hadron
ratios and multi-strange baryon production yet, though a good agreement between the
data and PYTHIA Perugia-2011 is
observed for charged kaons. In Pb--Pb
collisions the $\rm \bar{p}/\pi^{-}$ integrated ratio is significantly 
lower than statistical model predictions with a chemical freeze-out temperature $T_{ch} =
160-170$~MeV. The average transverse momenta and the spectral shapes indicate
a $\sim$10\% stronger radial flow than at RHIC energies. The comparison with
recent hydrodynamic calculations suggests that extra flow builds up in the
hadronic phase.


\begin{thebibliography}{99}
\bibitem{ref:ALICEperf}ALICE Collaboration,
  \Journal{J. Phys.}{G32}{1295}{2006}\\
  ALICE Collaboration, \Journal{J. Instrum.}{3}{S08002}{2008}
\vspace{-0.5mm}
\bibitem{ref:spectra900}ALICE Collaboration, \Journal{Eur. Phys. J.}{C71}{1655}{2011}
\vspace{-0.5mm}
\bibitem{ref:BarbaraSQM}B. Guerzoni (ALICE Collaboration), these proceedings
\vspace{-0.5mm}
\bibitem{ref:panos}ALICE Collaboration, \Journal{Phys. Rev. Lett.}{105}{072002}{2010}
\vspace{-0.5mm}
\bibitem{ref:AntoninSQM}A. Maire (ALICE Collaboration), these proceedings
\vspace{-0.5mm}
\bibitem{ref:DhevanSQM}D. Gangadharan (ALICE Collaboration), these proceedings
\vspace{-0.5mm}
\bibitem{ref:ALICEpbpb}ALICE Collaboration, \Journal{Phys. Rev. Lett.}{106}{032301}{2011}
\vspace{-0.5mm}
\bibitem{ref:blastwave}E. Schnedermann, J. Sollfrank and U. Heinz, \Journal{Phys. Rev.}{C48}{2462}{1993}
\vspace{-0.5mm}
\bibitem{ref:RHIC}STAR Collaboration, \Journal{Phys. Rev.}{C79}{034909}{2009}\\
  PHENIX Collaboration, \Journal{Phys. Rev.}{C69}{034909}{2004}\\
  BRAHMS Collaboration, \Journal{Phys. Rev.}{C72}{014908}{2005}
  \vspace{-0.5mm}
\bibitem{ref:statmodels}J. Cleymans \emph{et al.}, \Journal{Phys. Rev.}{C74}{034903}{2006}\\
  A. Andronic \emph{et al.}, \Journal{Phys. Lett.}{B673}{142}{2009}
\vspace{-0.5mm}
\bibitem{ref:MicheleQM}M. Floris (ALICE Collaboration), Quark Matter 2011, {\tt hep-ex/1108.3257}
\vspace{-0.5mm}
\bibitem{ref:purehydro}C. Shen, U. Heinz, P. Huovinen and H. Song, {\tt nucl-th/1105.3226}
\vspace{-0.5mm}
\bibitem{ref:SnellingsQM}R. Snellings (ALICE Collaboration), Quark Matter 2011
\vspace{-0.5mm}
\bibitem{ref:vishnu}U. Heinz, C. Shen, and H. Song, {\tt nucl-th/1108.5323}
\vspace{-0.5mm}
\bibitem{ref:vishnuspectra}U. Heinz and H. Song, private communication
\vspace{-0.5mm}
\bibitem{ref:NoferoSQM}F. Noferini (ALICE Collaboration), these proceedings
\vspace{-0.5mm}
\bibitem{ref:KalweitSQM}A. Kalweit (ALICE Collaboration), these proceedings
\vspace{-0.5mm}
\bibitem{ref:MariellaSQM}M. Nicassio (ALICE Collaboration), these proceedings
\end{thebibliography}
\end{document}